\documentclass[conference]{IEEEtran}
\IEEEoverridecommandlockouts

\usepackage{cite}
\usepackage{extarrows}
\usepackage{caption}
\usepackage{booktabs}  
\usepackage{amsmath,amssymb,amsfonts}
\usepackage{amsthm}
\usepackage{algorithmic}
\usepackage{graphicx}
\usepackage{textcomp}
\usepackage{xcolor}
\usepackage{url}
\def\BibTeX{{\rm B\kern-.05em{\sc i\kern-.025em b}\kern-.08em
    T\kern-.1667em\lower.7ex\hbox{E}\kern-.125emX}}
\begin{document}

\title{GenoHoption: Bridging Gene Network Graphs and Single-Cell Foundation Models\\
}

\author{\IEEEauthorblockN{1\textsuperscript{st} Jiabei Cheng}
\IEEEauthorblockA{\textit{Department of Automation} \\ \textit{School of Electronic Information and Electrical Engineering} \\ \textit{Shanghai Jiaotong University}\\
Shanghai, China \\
jiabei\_cheng@sjtu.edu.cn}
\and
\IEEEauthorblockN{2\textsuperscript{nd} Jiachen Li}
\IEEEauthorblockA{\textit{Department of Automation} \\ \textit{School of Electronic Information and Electrical Engineering} \\ \textit{Shanghai Jiaotong University}\\
Shanghai, China \\
lijc0804@sjtu.edu.cn}
\and
\IEEEauthorblockN{3\textsuperscript{rd} Kaiyuan Yang}
\IEEEauthorblockA{\textit{Department of Automation} \\ \textit{School of Electronic Information and Electrical Engineering} \\ \textit{Shanghai Jiaotong University}\\
Shanghai, China \\
yangkaiyuan@sjtu.edu.cn}
\and
\IEEEauthorblockN{4\textsuperscript{th} Hongbin Shen}
\IEEEauthorblockA{\textit{Department of Automation} \\ \textit{School of Electronic Information and Electrical Engineering} \\ \textit{Shanghai Jiaotong University}\\
Shanghai, China \\
hbshen@sjtu.edu.cn}
\and
\IEEEauthorblockN{5\textsuperscript{th} Ye Yuan *}
\IEEEauthorblockA{\textit{Key Laboratory of Biopharmaceutical Preparation and Delivery}\\ \textit{Chinese Academy of Sciences}\\
Beijing, China \\
\textit{State Key Laboratory of Biochemical Engineering, Institute of Process Engineering}\\
\textit{Chinese Academy of Sciences}\\
Beijing, China \\
yyuan@ipe.ac.cn}
}

\maketitle

\begin{abstract}
The remarkable success of foundation models has sparked growing interest in their application to single-cell biology. Models like Geneformer and scGPT promise to serve as versatile tools in this specialized field. However, representing a cell as a sequence of genes remains an open question since the order of genes is interchangeable. Injecting the gene network graph offers gene relative positions and compact data representation but poses a dilemma: limited receptive fields without in-layer message passing or parameter explosion with message passing in each layer. To pave the way forward, we propose GenoHoption, a new computational framework for single-cell sequencing data that effortlessly combines the strengths of these foundation models with explicit relationships in gene networks. We also introduce a constraint that lightens the model by focusing on learning the predefined graph structure while ensuring further hops are deducted to expand the receptive field. Empirical studies show that our model improves by an average of 1.27\% on cell-type annotation and 3.86\% on perturbation prediction. Furthermore, our method significantly decreases computational overhead and exhibits few-shot potential. Overall, GenoHoption can function as an efficient and expressive bridge, connecting existing single-cell foundation models to gene network graphs.\footnote{Github:\url{https://github.com/Bunnybeibei/GenoHoption}}
\end{abstract}

\section{Introduction}
Recently, foundation models, pre-trained on massive unlabeled datasets and fine-tuned on specific tasks, have made transformative and rapid progress in diverse applications. Single-cell biology, especially single-cell transcriptomics, is expected to benefit from the cross-task generalization ability of these large-scale models. Emerging high-quality models in this field such as scBERT~\cite{yang_scbert_2022}, scGPT~\cite{cui_scgpt_2024}, Geneformer~\cite{theodoris_transfer_2023}, scFoundation~\cite{hao_large_2023}, scSimilarity~\cite{heimberg_scalable_2023} and GeneCompass~\cite{yang_genecompass_2023} mainly employ a transformer-inspired architecture~\cite{vaswani_attention_2023}, viewing each cell as a ``sentence'' of genes.
\begin{figure}
  \centering
  \includegraphics[width=0.5\textwidth]{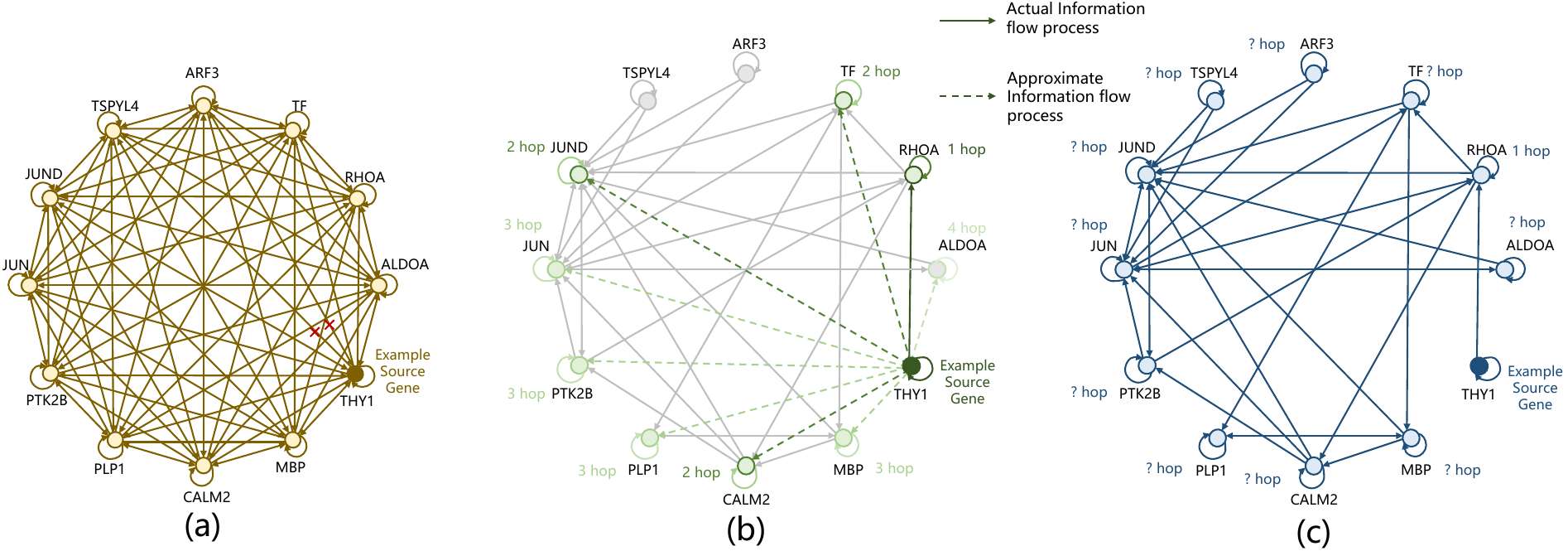}
    \caption{The illustration depicts the connections among twelve sample genes within a cell. (a) The framework of current single-cell foundation models. (b) The framework of our model. (c) The framework of prior gene network knowledge.}\label{intro}
\end{figure}
Given that the gene order within a cell can be swapped, various models represent this differently: for instance, scGPT~\cite{cui_scgpt_2024} evenly bins genes based on expression to introduce position embedding. At the same time, Geneformer~\cite{theodoris_transfer_2023} ranks genes according to their expression relative to the remaining genes in the cell. However, controversy remains about the most suitable way of organizing these varied genes in single-cell data. Additionally, considering the likelihood that any gene arrangement alteration could impact the results, the graph format may be more straightforward and robust as gene network graphs naturally offer gene relative positions~\cite{lazaros_graph_2023}.

The high-dimensional characteristic of single-cell sequencing data also challenges the regular use of self-attention~\cite{vaswani_attention_2023} within limited computational resources. Recent research employs efficient frameworks that originated to handle long text but raise unique issues when applied to single-cell data. For example, the memory-efficient approach utilized by scBERT~\cite{yang_scbert_2022} -- known as the Performer~\cite{choromanski_rethinking_2020} -- approximates the attention calculation, may fail to discern disparities at the individual gene level~\cite{nguyen_hyenadna_2023}, such as variations triggered by CRISPR knock-outs. Moreover, there remains a possible conflict between the high-density gene interconnections in these frameworks (Fig.\ref{intro}a) and the sparse direct gene connections in reality (Fig.\ref{intro}c). Considering these issues, the sparse-based efficient methods might be more appropriate, but they lack the gene-specific inductive bias like local windows for text sparsity. Thus, gene network graphs can also offer a solution to these complications.

The previous discussion suggests that gene network graphs could be valuable owing to their precise relative positions and reasonably sparse characteristics. It is also important to harness the strengths of pre-trained single-cell foundational models due to the incompleteness of prior network knowledge documented in databases. A possible strategy is creating links between adjacency matrices and attention matrices. However, this method could either significantly increase the learnable parameters, similar to equipping the Graph Attention Transformer~\cite{choromanski_graph_2021} in each layer, or result in a highly restricted receptive field, much like implementing an attention mask directly in each layer. Both scenarios oppose the original intention of integrating gene network graphs. We propose a novel computational framework for single-cell sequencing data entitled GenoHoption to address this dilemma. This framework aims to lighten and optimize the representations of single-cell foundational models by utilizing gene network knowledge. Our model consists of three parts: (1) Convert the single-cell gene sequence into the single-cell gene graph based on genetic and regulatory relationships. (2) Impose a decay constraint (Fig.\ref{intro}b) that only learns 1-hop parameters and deduces further hops through iteration, implemented by graph diffusion. (3) Detach node embeddings and restore them to the single-cell gene sequence for subsequent calculations. Our general framework can integrate various single-cell foundational models as backbones and improve their performance, as demonstrated in our experimental results. The contributions of this paper are as follows:
\begin{itemize}
\item We recognize the potential of gene structure patterns in mitigating the issues of position confusion and overwhelming computational resources encountered in current single-cell foundation models. This is the first work incorporating gene network graphs into these foundation models.
\item We apply graph diffusion to introduce decay constraints, efficiently injecting the gene network graphs and preserving the expressiveness of these pre-trained large-scale models.
\item Empirical experiments show that our model consistently and significantly outperforms various existing single-cell foundation models as backbones. More specifically, it achieves an average improvement of 1.27\% on cell-type annotation and 3.86\% on perturbation prediction. Upon further analysis of our model's performance on single-cell data, we see it exhibits potential for few-shot learning. It showcases a competitive, or even superior, memory efficiency compared to state-of-the-art frameworks tailored for long text.

\end{itemize}
\begin{figure*}[!t]
    \centering
	\includegraphics[width=1.0\textwidth]{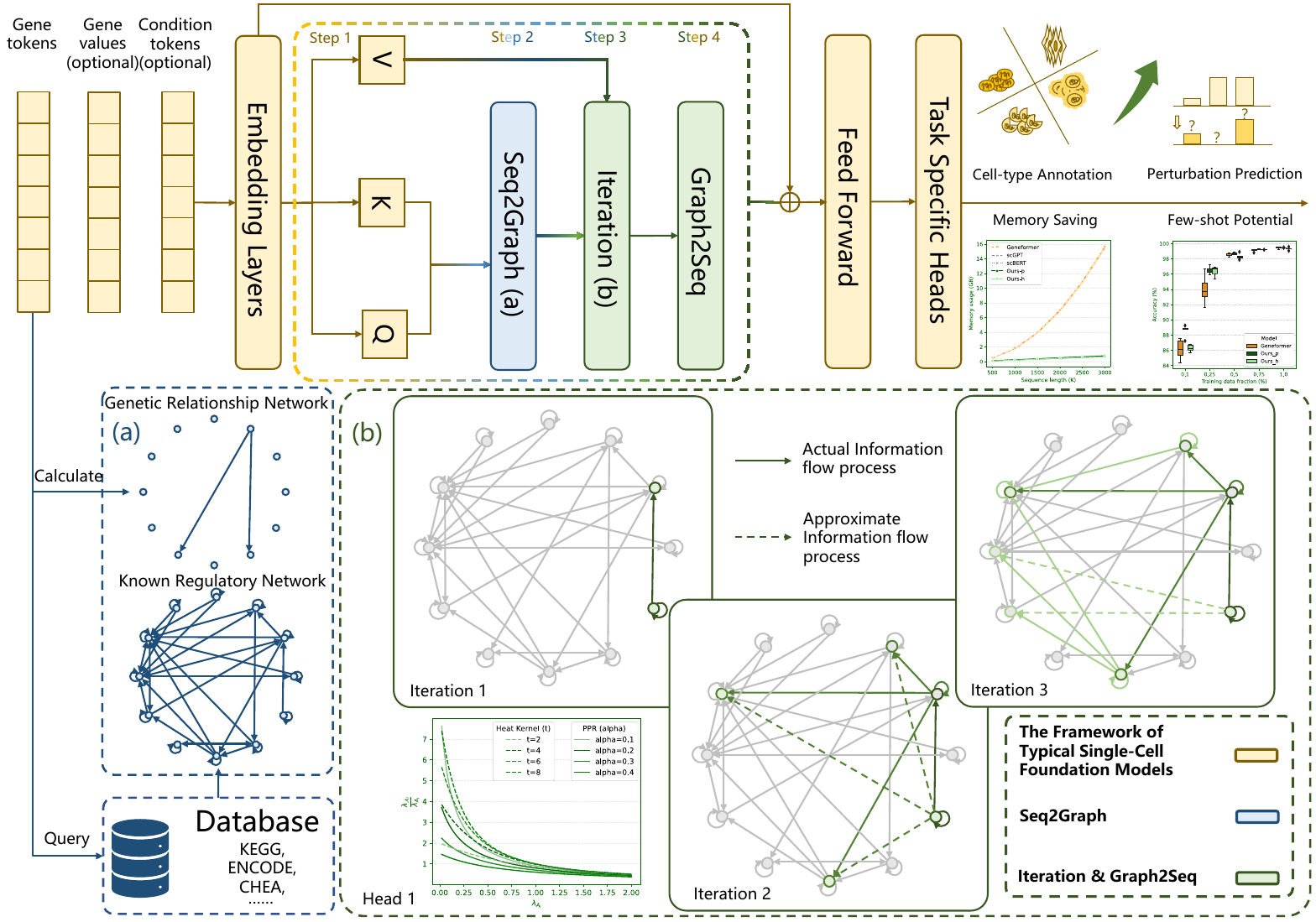}
	\caption{The layer architecture of our model. The attention computation consists of four steps: (1) Select a single-cell foundation model (in yellow) as the backbone for computing $QKV$ (Eq.\ref{QKV}). (2) Seq2Graph (in blue) transforms the single-cell gene sequences into single-cell gene network graphs. (3) Iteration (in green) broadens the receptive field of each gene without intense computation. (4) Graph2Seq (also in green) reverts the single-cell gene graphs into single-cell gene sequences for subsequent computations. Detailed data are available in the experiments section.
}
    \label{overview}
\end{figure*}
\section{Related Work}
\subsection{Single-Cell Foundation Models}
There is an emerging line of foundation models regarding single-cell data, which typically treat a cell as a sentence composed of genes but diverge in arranging these genes. Models like scBERT~\cite{yang_scbert_2022} and xTrimoGene~\cite{gong_xtrimogene_2023} keep the genes in their original order from the dataset without specific position encoding. scGPT~\cite{cui_scgpt_2024} bins genes according to their expression, assuming an even gene distribution across each bin, and uses bin numbers as position indicators. Geneformer~\cite{theodoris_transfer_2023} calculates a ranked list in which the position of a gene indicates its expression level relative to the remaining genes in the cell. At the same time, the sequence format makes it inconvenient to add domain structure knowledge, like gene regulatory networks and gene co-expression networks. GeneCompass~\cite{hao_large_2023} applies gene2vec~\cite{du_gene2vec_2019} to transform such knowledge into sequence format, but might still lack explicitness. To our knowledge, our work is a pioneering attempt that merges the gene network graph to tackle these unique problems in applying foundation models to single-cell biology.
\subsection{Efficient Transformers}
Given the computational complexity caused by the high-dimensional characteristics of single-cell data, several efficient transformer frameworks have been adopted in recent studies. For instance, scBERT~\cite{yang_scbert_2022} and xTrimoGene~\cite{gong_xtrimogene_2023} utilize Performer~\cite{choromanski_rethinking_2020}, a model approximates attention calculation using kernel-based method. Additionally, xTrimoGene employs an asymmetric encoding and decoding technique to enhance efficiency further. scGPT~\cite{cui_scgpt_2024} is built on FlashAttention~\cite{dao_flashattention_2022}, a memory-efficient framework that fully exploits speed differences across various hardware. On the other hand, another important line of work, the sparse-based methods, has yet to be developed, mainly because the text-tailored sparse feature does not suit interchangeable genes. Longformer~\cite{beltagy_longformer_2020} uses a combination of window attention and global attention, while Bigbird~\cite{zaheer_big_2020} and Diffuser~\cite{feng_diffuser_2022} reduce the window size and add random attention, with the latter further leveraging Personalized PageRank~\cite{wei_topppr_2018} for message passing. Inspired by these methods, our model propels the gene network graph to sparsely optimize existing single-cell foundation models.
\section{Methods}
\subsection{Problem Formulation}
It is advantageous to utilize both gene network graphs for explicit gene positions and sparsification, as well as the capabilities of large-scale single-cell models. We first formulate the dilemma between efficiency and expressiveness encountered in pursuing this objective. All random variables and their corresponding realizations are denoted by bold and thin letters.

\paragraph{Two perspectives of Transformer.}Transformers and self-attention mechanism ~\cite{vaswani_attention_2023} was introduced for sequence modelling. The input sequence, with $n$ tokens, can be denoted as $\pmb{x} = [x_1, x_2, ..., x_n]$, where $\pmb{x}\in R^{n\times d}$ and each token $x_i$ is a $d$-dimensional vector. By using linear projections: 
\begin{equation}\label{QKV}
Q = \pmb{x}W_Q, \quad K = \pmb{x}W_K, \quad V = \pmb{x}W_V.
\end{equation}

the attention mechanism maps $\pmb{x}$ to ``query'' ($Q$), ``key'' ($K$), and ``value'' ($V$) spaces. Subsequently, it calculates the attention matrix $A$ as the scaled dot-product of $Q$ and $K$: 
\begin{equation}
\label{attention}
    A = softmax(\frac{QK^{T}}{\sqrt{d}}),
\end{equation}
and finally updates token values by multiplying $A$ with the value matrix $V$ :
\begin{equation}
\label{gather}
Attn(\pmb{x}) = AV.
\end{equation}

When treating the attention as edges in a graph and adopting message passing on these edges, another perspective of Transformer could be obvious. Assume a fully connected graph, denoted as \pmb{$G$}, with $n$ nodes represented as $\pmb{x}=[x_1, x_2, ..., x_n]$. The attention mechanism can be performed similarly. Specifically, it can be decomposed into two steps: (1) Message computation: This step involves calculating the message sent from node $x_j$ to $x_i$, as represented by $A_{i,j}$ in Eq.\ref{attention}. (2) Message aggregation: This step is analogous to Eq.\ref{gather} where $Attn(x_i)$ represents the aggregation of the incoming values to node $x_i$.

\paragraph{The Dilemma between Efficiency and Expressiveness.}When a fully connected graph turns to a sparse and directed one, it also gains the potential for sparsification by attending to neighbours instead of all tokens. This approach is particularly suitable for single-cell sequencing data with long and certain structured patterns, such as gene networks. By treating genes $\pmb{x}$ as nodes $\mathcal{V}$ and employing the gene network to create an adjacency matrix $A$, we can generate a single-cell gene graph $\pmb{G}=(\mathcal{V},\mathcal{E})$ to replace the original sequence $\pmb{x}$ during attention computation. Each directed edge within the graph corresponds to a query-key pair. In this way, we can easily leverage the parameters of the pre-trained single-cell foundation model and inject the graph structure. The resulting sparse self-attention mechanism, as described in Eq.\ref{attention}, can be rewritten in a token-wise form as
\begin{equation}
\label{sparse}
Attn(x_i) = softmax(\frac{Q_{i}K^{T}_{Ne(i)}}{\sqrt{d}})V_{Ne(i)},
\end{equation}

where $x_i$ is the $i$-th input gene to update value and $Ne(i)$ denotes the neighbors of gene $i$ in the gene network graph \pmb{$G$}. Sparse attention could inevitably be lossy due to 1-hop messaging in each layer, which necessitates more layers and dilutes the advantage of fewer parameters. In this scenario, there is a necessary trade-off between the efficiency of training and the expressiveness of the model.
\subsection{Model Formulation}\label{model formula}
 An alternative could be to increase the per-layer receptive field predefined by $\pmb{G}$ without increasing the total layer count. For example, we can adopt the graph diffusion mechanism to get multi-hop attention scores: 
\begin{equation}
\label{diffuse}
    \mathcal{A} = \sum_{k=0}^{\infty}\theta_{k}A^{k},
\end{equation}
where $A$ is the adjacency matrix or computed sparse attention matrix, and the weighting coefficient $\theta_k$ satisfies $\sum_{k=0}^{\infty}\theta_k = 1, \theta_k \in [0,1]$. $\theta_k$ can be seen as a specific constraint that simplifies the model by prioritizing the learning of $1$-hop parameters $A$ while also ensuring the deduction of $k$-th hop.

We adopt two varieties of this coefficient: Personalized PageRank (PPR) ~\cite{wei_topppr_2018}, labelled as \textbf{GenoHoption-p} and specified by
\begin{equation}
\label{ppr}
    \theta_{k} = \alpha(1-\alpha)^{k},
\end{equation}
with the teleport probability $\alpha$, and Heat Kernel-based Diffusion ~\cite{kloster_heat_2014}, labeled as \textbf{GenoHoption-h} and specified by 
\begin{equation}
\label{heat}
    \theta_{k} = e^{-t}\cdot\frac{t^{k}}{k!}, 
\end{equation} with $t$.

Given that Eq.\ref{diffuse} involves the power calculation of attention matrices -- an operation that could be highly costly in computation for long sequences, even factoring in the sparsity -- we implement it in a power iteration way. This technique allows us to reach linear computational complexity by approximating the first $K$ diffusion steps. Detailed power iteration steps are shown in the following propositions.

\paragraph{Proposition 1.} \emph{Assume $V_{k}$ is the updated value after $k$ steps, if the iteration of \textbf{GeneHoption-p} is defined as 
\begin{align}\label{ppr_iter}
\begin{cases}
V_{0} = V\\
V_{k}=(1-\alpha)AV_{k-1}+\alpha V,
\end{cases}
\end{align}
then $\lim_{k\to \infty} V_{k}=\mathcal{A}V$.}

\begin{proof}
Let $K \textgreater 0$ be the total number of iterations. After $K$-th iteration, we can get
\begin{align}
V_{K}&=((1-\alpha)^{K}A^{K}+\alpha\sum^{K-1}_{i=0}(1-\alpha)^{i}A^{i})V \nonumber \\&=(\sum_{k=0}^{K-1}\alpha(1-\alpha)^{k}A^{k})V \nonumber\\&\xlongequal{K\to \infty}\mathcal{A}V.
\end{align}The term $(1-\alpha)^{K}A^{K}$ converges to $0$ as $\alpha \in (0,1]$ and each $A^{K}_{i,j}\in(0,1]$ when $K \to \infty$. $\mathcal{A}$ denotes PPR ~\cite{wei_topppr_2018} based diffusion matrix.
\end{proof}
\paragraph{Proposition 2.} \emph{Assume $V_{k}$ is the updated value at $k$-th step, if the iteration of \textbf{GeneHoption-h} is defined as 
\begin{align}\label{heat_iter}
\begin{cases}
V_{0} = e^{-t}V\\
V_{k}= \dfrac{tA}{k}V_{k-1}\\
\end{cases}
\end{align}then $\lim_{K\to \infty} \sum_{k=0}^{K}V_{k}=\mathcal{A}V$.}
\begin{proof}
Let $K \textgreater 0$ be the total number of iterations. After $K$-th iteration, we can sum them as
\begin{align}
\sum_{k=0}^{K}V_{(k)}&=(V_0+V_1+V_2+...+V_K)
\nonumber\\&=(I+tA+\dfrac{t^{2}A^{2}}{2}+...+\dfrac{t^{K}A^{K}}{K!})e^{-t}V \nonumber\\&=(\sum^{K}_{k=0}e^{-t}\dfrac{t^{k}}{k!}\cdot A^{k})V\nonumber\\&\xlongequal{K\to \infty}\mathcal{A}V,
\end{align} where $\mathcal{A}$ denotes Heat Kernel ~\cite{kloster_heat_2014} based diffusion matrix.
\end{proof} 
This way, we can avoid the direct computation of attention matrix powers while obtaining a sufficient in-layer receptive field.
\begin{table*}[!t]
    \centering 
    \begin{tabular}{lcccccc}
   
        \toprule
        Datasets & \multicolumn{2}{c}{\textbf{Myeloid.}} & \multicolumn{2}{c}{\textbf{Pancreas.}} & \multicolumn{2}{c}{\textbf{MS.}}\\
        \cmidrule(lr){2-3}\cmidrule(lr){4-5}\cmidrule(lr){6-7}
        Metrics & \textbf{Accuracy} & \textbf{F1-score} & \textbf{Accuracy} & \textbf{F1-score} & \textbf{Accuracy} & \textbf{F1-score}\\
        \midrule
        \textbf{scBERT}     & $61.55\pm4.60$ & $35.11\pm1.82$& $95.61\pm1.29$ & $65.44\pm5.42$& $74.55\pm7.22$ & $57.37\pm7.36$  \\
        \hline
                \textbf{Geneformer}     & $78.30\pm1.65$ & $48.33\pm3.82$ &$99.13\pm0.18$& $71.34\pm3.18$ &$55.42\pm1.77$  &  $32.98\pm2.81$  \\
        \textbf{Geneformer + Longformer} &$64.21\pm1.15$ & $40.43\pm 2.91$ & $96.04\pm1.39$ &$66.91\pm8.92$&$60.61\pm0.75$  & $41.17\pm5.60$ \\
        \textbf{Geneformer + BigBird} & $58.74\pm2.71$& $25.76\pm5.33$  & $91.04\pm1.28$ &$50.37\pm3.27$& $46.25\pm0.77$ & $22.04\pm1.80$\\
        \textbf{Geneformer + Diffuser} & $67.72\pm0.74$ & $47.62\pm1.58$ & $96.59\pm0.05$ & $69.71\pm4.42$& $59.11\pm1.19$ &  $37.70\pm2.47$ \\
        \textbf{Geneformer + ours-h}& $\underline{81.77\pm0.12}$ & $\underline{48.96\pm2.17}$  & $\pmb{99.25\pm0.04}$ & $\underline{74.51\pm2.41}$&   $\underline{75.02\pm0.93}$ &  $\underline{56.18\pm2.82}$  \\
        \textbf{Geneformer + ours-p}&\pmb{$82.31\pm0.32$} &\pmb{$50.24\pm2.07$}  & \underline{$99.24\pm0.06$} & \pmb{$74.62\pm1.22$}& $\pmb{80.21\pm0.58}$ &  $\pmb{63.56\pm0.60}$ \\
        \hline
         \textbf{scGPT}     & $62.43\pm 1.50$ & $36.43\pm1.93$& $95.57\pm1.67$ & $68.93\pm6.86$&  $84.35 \pm 2.48$ & $\underline{68.00 \pm 3.75}$     \\
        \textbf{scGPT + Longformer} & $59.36\pm 1.52$&$33.55\pm1.21$& $95.43\pm1.69$ & $67.06\pm7.80$&$83.82\pm0.66$  &$66.81\pm2.51$   \\
        \textbf{scGPT + BigBird}  & $62.43\pm2.92$&$35.69\pm1.55$& $96.46\pm0.95$ & $68.06\pm4.43$& $81.27\pm4.78$ & $62.94\pm4.75$\\
        \textbf{scGPT + Diffuser}  &  $58.82\pm1.79$ & $32.92\pm1.57$& $96.15\pm0.44$ & $62.33\pm2.21$&  $73.98 \pm 1.20$& $57.57 \pm 0.68$\\
        \textbf{scGPT + ours-p}  &$\underline{64.51\pm1.58}$  & $\pmb{37.03\pm2.00}$& \underline{$97.52\pm0.31$} & \underline{$69.23\pm5.68$}&  \underline{$84.43 \pm 1.07$} & $66.88 \pm 1.59$      \\
        \textbf{scGPT + ours-h}  & $\pmb{65.03\pm 1.28}$ & $\underline{36.46\pm0.69}$ & $\pmb{97.62\pm0.31}$ & $\pmb{69.74\pm1.95}$ & $\pmb{84.98 \pm 1.36}$ &  $\pmb{68.59 \pm 2.47}$  \\
        \bottomrule
    \end{tabular}
    \caption{Performance comparison with baselines on three single-cell type annotation datasets (referred to as Myeloid, Pancreas and MS.) in terms of Accuracy (\%) and F1-score (\%). The best and the runner-up results are highlighted in \textbf{bolded} and \underline{underlined}, respectively. We emphasize the comparison against ``* + Longformer'', ``* + BigBird'' and ``* + Diffuser'', as well as scBERT, which also functions as a proxy for Performer.}
    \label{table1}
\end{table*}
\begin{figure*}[!t]\centering 
	\includegraphics[width=1.0\textwidth]{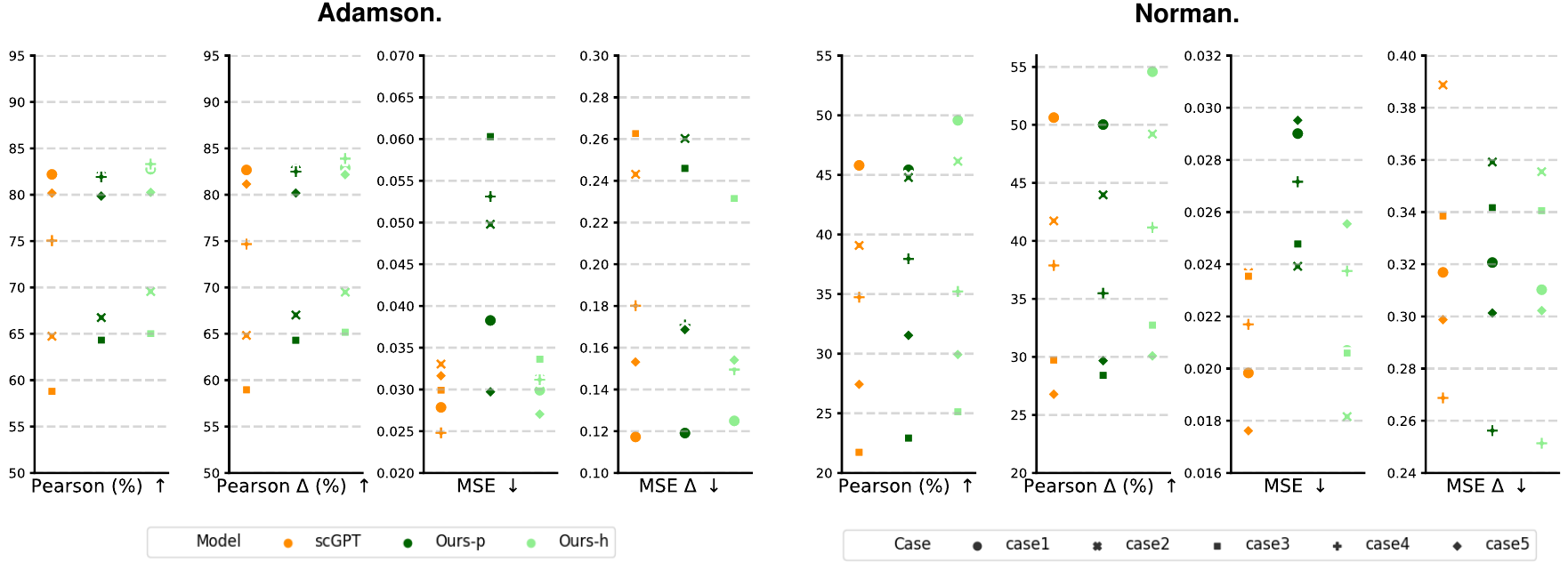}
	\caption{Performance comparison with baselines on two single-cell perturbation prediction datasets (referred to as Adamson. and Norman.) in terms of Pearson metrics (\%), Mean Square Error ($\times$100) on non-zero genes, and the top 20 differentially expressed genes ($\Delta$). 
Different colors represent different methods, and points of different shapes signify different cases of unseen gene perturbation. $\uparrow$ indicates that higher values are better and $\downarrow$ indicates that lower values are preferred.
}
    \label{pert}
\end{figure*}
\subsection{Model Instantiations}
\paragraph{Seq2Graph.}To simplify the process of converting single-cell gene sequences into single-cell gene graphs, we designed a module named \textbf{Seq2Graph} (Fig.\ref{overview}a). This module initially constructs a complete gene network graph using the training dataset and supports gene tokens to retrieve corresponding structures in every attention computation. The graph is composed of two parts:
\begin{enumerate}
    \item Co-expression patterns within the cell. These are generated by calculating the Pearson correlations $\rho(x_i,x_j)$ between genes $x_i$ and $x_j$. For each gene $x_i$, we connect it to the top $H$ genes, which have the highest $\rho_{x_i,x_j}$ and are above a specific threshold $\delta$.
    \item Gene regulation relationships. These are gleaned from 10 different datasets: KEGG~\cite{kanehisa_kegg_2000}, RegNetwork~\cite{liu_regnetwork_2015}, TRRUST~\cite{han_trrust_2015}, HTRIdb~\cite{bovolenta_htridb_2012}, EVEX~\cite{van_landeghem_exploring_2012}, JASPAR~\cite{mathelier_jaspar_2014}, CHEA~\cite{lachmann_chea_2010}, TRANSFAC~\cite{matys_transfac_2006}, MOTIFMAP~\cite{xie_motifmap_2009}, and ENCODE~\cite{noauthor_encode_2004}, similar to the operation of DGP-AMIO~\cite{yang_integration_2023}.
\end{enumerate}

\paragraph{Iteration \& Seq2Graph.}After obtaining gene network graphs \pmb{$G$}, where the nodes $\pmb{x}$ come from single-cell sequences and the edges are derived by \textbf{Seq2Graph}, we apply graph diffusion through a module called \textbf{Iteration} (Fig.\ref{overview}b). This module implements iterations shown in Eq.\ref{ppr_iter} and Eq.\ref{heat_iter}, replacing the traditional sequence-based dense attention calculation in each layer. This process is highly compatible and requires no additional parameters. It can be viewed as specifying a latent space subject to specific decay constraints involving multiple hops' relationships, thus necessitating the model to comply during training. After sufficient iterations, the node embeddings can be easily detached and reorganized according to the original sequences for subsequent calculations. This procedure is named \textbf{Seq2Graph}. The overview of our model is demonstrated in Fig.\ref{overview}.
\section{Experiments}
Experiments are conducted on 3 benchmark datasets for cell-type annotation and 2 for perturbation prediction. Each experiment is repeated 5 times, with the mean and standard deviation reported, running on a machine with two Intel Xeon Gold 6138 CPUs, four RTX 3090 GPUs, and 503GB of RAM.
\begin{figure*}[!t]\centering 
	\includegraphics[width=1.0\textwidth]{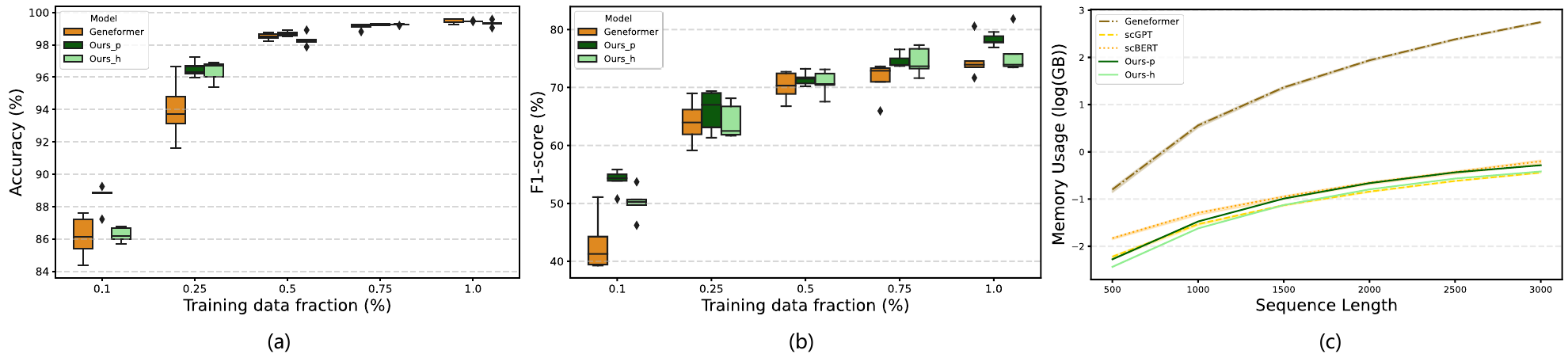}
	\caption{(a) Accuracy (\%) variation with different training set fractions on the test data.
(b) F1-score (\%) variation with different training set fractions on the test data. All experiments are conducted on Geneformer with the Pancreas dataset. (Note: Detailed data and the usage of scGPT as the backbone in the same setting can be found in the Appendix C.)
(c) Comparison of space complexity. The units are given in milliseconds (ms) and the logarithm of gigabytes (log(GB)), respectively. It reports the average and the standard deviation of the results repeated 10 times on a RTX 3090 GPU with a batch size of 32.
}
    \label{fewshot}
\end{figure*}
\subsection{Datasets and Setups}\label{setups}
\paragraph{Datasets and protocols.}The datasets for the cell-type annotation -- \textbf{Myeloid}~\cite{cheng_pan-cancer_2021}, and \textbf{Pancreas}~\cite{chen_transformer_2023}, \textbf{Multiple Sclerosis}~\cite{schirmer_neuronal_2019} -- which have already been split into training and testing datasets, are gathered from scGPT~\cite{cui_scgpt_2024}. We keep the test dataset unchanged and set aside 25\% of the training set as a validation split. We further split the datasets into four different fractions in the few-shot study. The other two datasets, \textbf{Adamson}~\cite{adamson_multiplexed_2016} and \textbf{Norman}~\cite{norman_exploring_2019}, are used for perturbation prediction and are generated from GEARS~\cite{roohani_predicting_2023}. Similar to the previous datasets, we allocate 75\% of the samples for training. Note that all test splits contain cell or perturbation types that differ from those in the training and validation sets. More details of datasets are provided in Appendix A.

\paragraph{Metrics.}We report both Accuracy and F1-score for the cell-type annotation for a comprehensive comparison. Additionally, for unseen perturbation prediction, we evaluate the Pearson correlation and Mean Square Loss on non-zero genes and the top 20 differentially expressed genes ($\Delta$).

\paragraph{Baselines.}Our method can theoretically be compatible with any existing single-cell foundation model to enhance performance. However, since some models are not open-source during this research, we adopt two backbones: \textbf{scGPT}~\cite{cui_scgpt_2024}, and \textbf{Geneformer}~\cite{theodoris_transfer_2023} into our method. In addition to these two foundation models, we compare our method to \textbf{scBERT}~\cite{yang_scbert_2022}, which also functions as a proxy for \textbf{Performer}~\cite{choromanski_rethinking_2020} -- a kernel-based memory-efficient method, as well as three sparse-based efficient transformers: \textbf{Longformer}~\cite{beltagy_longformer_2020}, \textbf{BigBird}~\cite{zaheer_big_2020}, and \textbf{Diffuser}~\cite{feng_diffuser_2022}. These methods all use the same parameters in their source code. Training details are specified in Appendix B.

\subsection{Performance Comparison}
\paragraph{Improvements to Cell-Type Annotation.}Table \ref{table1} shows that, in most cases, our method uses much fewer parameters but significantly improves the Accuracy and F1-score of the baselines (p\_values: $[0.0015, 0.1898, 3.26\times10^{-8}]$). In addition, our method outperforms other efficient variations such as ``+ Longformer~\cite{beltagy_longformer_2020}''. These improvements might be attributed to our method employing gene structure features, making it more appropriate for cell-type annotation than previous sparse-based methods, which mostly rely on local features. GenoHoption-p enhances Genefomer more significantly, while GenoHoption-h works better with scGPT. We speculate this difference may arise from the distinct manifolds pre-learned by these models, with each space being more readily adaptable to one of the decay constraints.
\paragraph{Improvements to Perturbation Prediction.}Fig \ref{pert} presents the comparison results for perturbation prediction. Since Geneformer does not support predicting specific perturbation values, our benchmarking is only against scGPT. Our methods consistently surpass its backbone in different unseen perturbed-gene cases. It is noticed that although scGPT has a lower mean square error metric on all genes, it performs worse in predicting differentially expressed gene values and Pearson metrics. We speculate this phenomenon is because scGPT uses a fully connected attention mechanism that fits much more noise than our methods.
\paragraph{Few-shot Potential.}We hypothesize that our method may exhibit more significant improvements in a few-shot setting due to its ability to narrow down the search space. This narrowing is achieved by using prior network knowledge to preclude specific gene interactions, which would otherwise require a large amount of training data for discovery, and directing models to a specific, decay-constrained domain rather than the entire potential space. As shown in Fig.\ref{fewshot}(a)(b), this hypothesis generally aligns with the results conducted on Geneformer~\cite{theodoris_transfer_2023} using various proportions of Pancreas ~\cite{chen_transformer_2023} datasets. The consistent results with scGPT~\cite{cui_scgpt_2024} can be found in Appendix C.
\begin{table*}[!t]
    \centering
    \begin{tabular}{lcccccc}
        \toprule
        Datasets & \multicolumn{2}{c}{\textbf{Myeloid.}} & \multicolumn{2}{c}{\textbf{Pancreas.}} & \multicolumn{2}{c}{\textbf{MS.}}\\
        \cmidrule(lr){2-3}\cmidrule(lr){4-5}\cmidrule(lr){6-7}
        Metrics & \textbf{Accuracy} & \textbf{F1-score}& \textbf{Accuracy} & \textbf{F1-score} & \textbf{Accuracy} & \textbf{F1-score}\\
        \midrule
        w/o GRN    & 
        \underline{$63.44\pm1.92$} &
        \underline{$35.86\pm2.26$}  & 
        $95.78\pm1.45$& 
        $63.86\pm5.00$&
        $81.48\pm2.06$ &  
        $63.36\pm1.69$ \\
        
        w/o Co    &  
        $63.36\pm1.43$ &
        $35.18\pm2.00$ & 
        \underline{$97.38\pm0.61$} & 
        \underline{$68.43\pm3.21$} &
        $82.70\pm2.53$ &  
        $63.24\pm2.43$ \\

        w/o iter   & 
        $62.15\pm2.06$ &
        $35.22\pm1.55$ &
        $95.80\pm0.62$ &
        $63.79\pm4.00$ &
        \underline{$84.02\pm0.81$} &
        \underline{$65.34\pm1.08$} \\

        \textbf{scGPT + ours-p}  &
        \pmb{$64.51\pm1.58$}  &
        $\pmb{37.03\pm2.00}$ &
        \pmb{$97.52\pm0.31$} &
        \pmb{$69.23\pm5.68$} &
        \pmb{$84.43 \pm 1.07$} &
        \pmb{$66.88 \pm 1.59$}\\

        \bottomrule
    \end{tabular}
    
    \caption{Ablation study performed on GenoHoption-p for cell-type annotation. The best and the runner-up results are highlighted in \textbf{bolded} and \underline{underlined}, respectively. }
    \label{ablation}
\end{table*}
\begin{figure*}[!t]
\centering
	\includegraphics[width=1.0\textwidth]{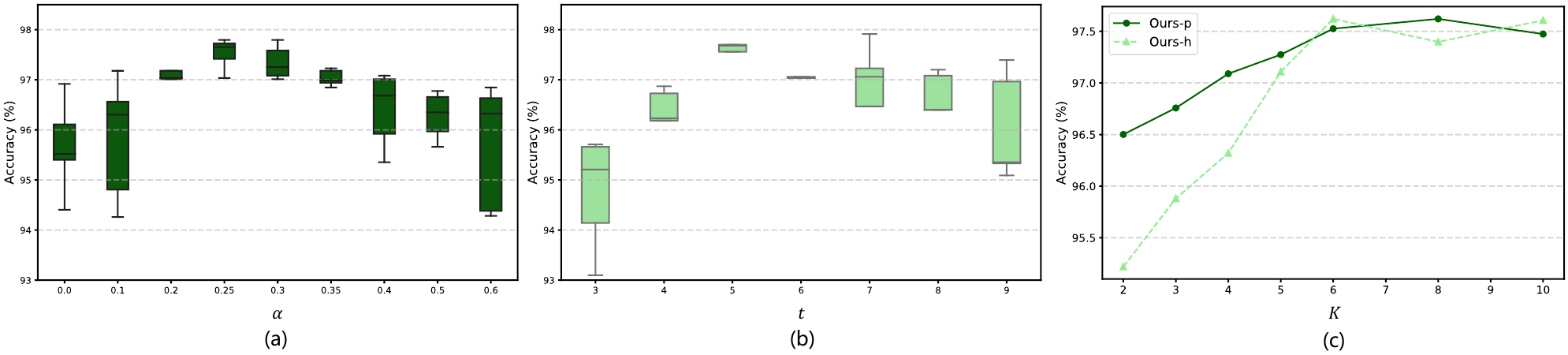}
	\caption{The impact of different hyper-parameters on the performance: 
 (a) Varying the specific $\alpha$ in Eq.\ref{ppr}. 
 (b) Varying the specific $t$ in Eq.\ref{heat}. 
 (c) Varying the specific iteration number $K$.(Note that (c) simply shows the average impact for the sake of clarity. The detailed data can be found in Appendix D.)}
    \label{hyper}
\end{figure*}
\subsection{Components and Hyper-parameter Study}
\paragraph{Memory Conservation.}We conducted 10 repetitions of a batch comprising 32 inputs with varying sequence lengths. As the Performer only supports processing one sequence per GPU, we used the ratio between the Performer and Standard Attention with a batch size of 1 to infer the performance of the Performer with a batch size of 32. Fig.\ref{fewshot}(c) presents a comparison of GPU memory usage among different models, showing that our method outperforms most, except scGPT~\cite{cui_scgpt_2024} which utilizes FlashAttention~\cite{dao_flashattention_2022}. However, the current version of FlashAttention does not yet support arbitrary attention masks, making it impractical to incorporate gene networks through the commonality between the attention matrix and the adjacency matrix. Thus, our method is a more viable framework than FlashAttention for single-cell foundation models since it produces an explicit adjacency matrix and consumes deficient memory. Although using smaller windows with Longformer~\cite{beltagy_longformer_2020} and Bigbird~\cite{zaheer_big_2020} could also decrease memory usage, it may not make sense when dealing with interchangeable genes within a cell. In addition, our method saves a considerable amount of memory compared to the Graph Attention Transformer (GAT)~\cite{choromanski_graph_2021} with the equivalent in-layer receptive field. This highlights its potential as an efficient framework for emerging single-cell foundational models.
\paragraph{Ablation Study.}In this section, we analyze the contributions of various model components to the final performance. Table \ref{ablation} provides detailed ablation experimental results for the cell-type annotation. The abbreviation ``w/o GRN'' indicates the absence of gene regulation network knowledge from the databases, ``w/o Co'' signifies the exclusion of genetic structure patterns from the training dataset, and ``w/o iter'' is analogous to directly using the attention mask. In general, although components have varying degrees of impact on different datasets, we recommend including all of them for the best performance.

\paragraph{Hyper-parameter Sensitivity Study.}We investigate the sensitivity of our method to these three hyper-parameters on the Pancreas~\cite{chen_transformer_2023} dataset: the $\alpha$ in Eq.\ref{ppr}, the $t$ in Eq.\ref{heat} and the specific number of iterations $K$. Fig.\ref{hyper}(a) illustrates the performance relevant to different values of $\alpha$. It shows that the performance of our method degrades when $\alpha$ is either too small (e.g., $\alpha = 0, 0.1$) or too large (e.g., $\alpha = 0.4, 0.5$). This could be explained that an overly weak decay could introduce more confusion, or an overly strong decay might restrict the receptive field. This trend is also observed in the performance regarding different $t$ values in Fig.\ref{hyper}(b). Fig.\ref{hyper}(c) presents the performance of our method by varying the number of iterations $K$. A significant improvement with increasing $ K $, but it gradually saturates when $K \geq 6 $, with some fluctuations
\section{Conclusion}
In this paper, we propose a general framework that can incorporate existing single-cell foundation models as backbones to bridge their perceived shortcomings in genetic structure. Specifically, we manage to transform gene sequences into gene network graphs for explicit gene relative positions and efficient sparsification. We also preserve models’ expressiveness through decay constraints imposed by the graph diffusion. Extensive experimental results demonstrate that our model yields consistent and significant improvements and exhibits the few-shot potential. Furthermore, regarding single-cell sequencing data, our model showcases competitive or even superior memory efficiency compared to state-of-the-art efficient transformers tailored for long text. We envision our model as a practical framework for budding single-cell foundation models.
\clearpage
\bibliography{GenoHoption}
\clearpage
\appendix
\begin{table}[!t]
    \centering
    \begin{tabular}{lrr}
        \toprule
        Datasets & \# Tokens & \# Samples\\
        \midrule
        Multiple Sclerosis.~\cite{schirmer_neuronal_2019}    &  $2808$ &  $7844 + 13468$  \\
        Myeloid.~\cite{cheng_pan-cancer_2021}    &  $2809$  &  $9748 + 3430$ \\
        Pancreas.~\cite{chen_transformer_2023}   & $2999$  &  $10600 + 4218$  \\
        Adamson.~\cite{adamson_multiplexed_2016}  &  $4399$  & $27573 + 5377$  \\
        Norman.~\cite{norman_exploring_2019}  & $4547$  & $28556 + 16123$  \\
        \bottomrule
    \end{tabular}
    
    \caption{Data statistics of used datasets. Within the ``\# Samples'' column, the value before the ``$+$'' represents the total count of training and validation samples, and the value following ``$+$'' refers to the number of test samples.}
    \label{dataset}\end{table}
\begin{table}[!t]
    \centering
    \begin{tabular}{lr}
        \toprule
        Parameters & Values\\
        \midrule
        The $\alpha$ of GenoHoption-p   &  $0.25$ \\
        The $t$ of GenoHoption-h  &  $5.0$ \\
        Iteration steps $K$   & $6$\\
        The mean degrees of MS./Myeloid./Pancreas.  &  $13/22/33$   \\
        Top gene number $H$. & $20$ \\
        Pearson threshold $\sigma$ &$0.4$\\
        \bottomrule
    \end{tabular}
    
    \caption{Hyperparameter Settings. Only the modified or newly added hyper-parameters are displayed. Parameters not shown remain consistent with those reported in the original paper.}
    \label{parmas}
\end{table}
\begin{figure*}[!b]\centering 
	\includegraphics[width=1.0\textwidth]{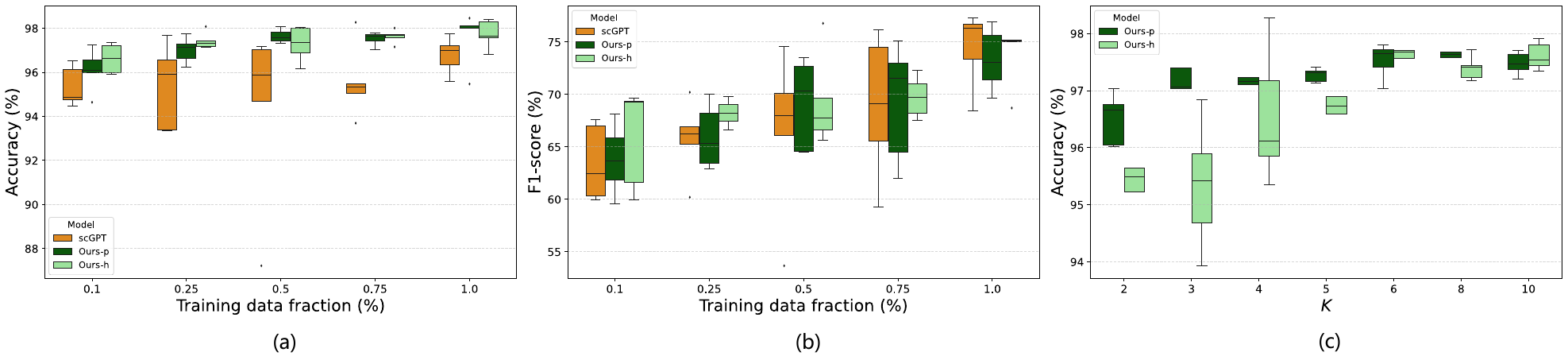}
	\caption{(a) Accuracy(\%) variation with different training set fractions on the test data.
(b) F1-score(\%) variation with different training set fractions on the test data. All experiments are conducted on scGPT with the Pancreas dataset. (c) Detailed data in varying iteration steps $K$ for each version of our model.}
    \label{fewshot_scgpt}
\end{figure*}
\section{Dataset Overview}
For cell-type annotation, we use 3 datasets, namely Multiple Sclerosis ~\cite{schirmer_neuronal_2019}, Myeloid ~\cite{cheng_pan-cancer_2021}, and Pancreas~\cite{chen_transformer_2023}. We use 2 datasets for perturbation prediction: Adamson~\cite{adamson_multiplexed_2016} and Norman~\cite{norman_exploring_2019}. Please see Table \ref{dataset} for relevant statistics, and the following are the introductions to these datasets.
\subsection{Cell-Type Annotation Dataset Introduction}
\paragraph{Multiple Sclerosis.}The multiple sclerosis~\cite{schirmer_neuronal_2019} dataset consists of neuronal cell types collected from healthy donors and patients with multiple sclerosis. The pre-processed data we utilize is downloaded from the authors of scGPT~\cite{cui_scgpt_2024}, adhering to their pre-processing and train/test splits methods. The division between the train/test for this dataset mimics a dataset shift, including cells from $9$ healthy donors for the training set and cells from $12$ patients with multiple sclerosis for the test set. The actual cell-type labels are derived from the original publication of the data. 

\paragraph{Myeloid.}The Myeloid dataset~\cite{cheng_pan-cancer_2021} includes tumour-infiltrating myeloid cells from various tumour types. We utilize the pre-processed data provided by the authors of scGPT~\cite{cui_scgpt_2024}, following the same pre-processing procedures and train/test divisions that they do. More specifically, the dataset includes cells from $9$ distinct cancer types, with the train set comprised of cells from UCEC, PAAD, THCA, LYM, cDC2, and KIDNEY cancer types, while the test set comprises cells from MYE, OV-FTC, and ESCA.
\paragraph{Pancreas.}The pancreas dataset~\cite{chen_transformer_2023} includes human pancreas cells from five separate single-cell RNA sequencing studies reprocessed by Chen and her team. We download and use the pre-processed data as provided by the authors of scGPT, thus keeping the same pre-processing steps and train/test distributions as they do. The training set contains $13$ cell types (alpha, beta, ductal, acinar, delta, PSC, PP, endothelial, macrophage, mast, epsilon, Schwann, and T-cell). It is gathered from two datasets, whereas the test set includes $11$ different cell types (alpha, beta, ductal, PP, acinar, delta, PSC, endothelial, epsilon, mast, MHC class II) collected from the remaining three datasets. 
\subsection{Perturbation Prediction Dataset Introduction}
\paragraph{Adamson.}This dataset comprises gene expression data extracted from the K562 leukaemia cell line perturbed by Pertub-seq~\cite{adamson_multiplexed_2016}. We use the pre-processed data provided by the authors of scGPT~\cite{cui_scgpt_2024}, following the same pre-processing protocols they used along with their train/test divisions. It includes $87$ unique one-gene perturbations, each replicated in approximately a hundred cells.
\paragraph{Norman.}The Norman perturbation dataset likewise includes gene expression data from the K562 leukaemia cell line, perturbed by Pertub-seq~\cite{norman_exploring_2019}. We employ the pre-processed data supplied by the authors of scGPT~\cite{cui_scgpt_2024}, observing the identical pre-processing steps and train/test divisions. This dataset has $131$ disturbances involving two genes and $105$ disturbances to one gene; each disturbance is replicated between three to seven hundred cells.
\section{Experiment Details}\label{experiment details}
We use all default settings and hyper-parameters provided in the source code of every single-cell foundation model, including a learning rate of $1 \times 10^{-4}$, batch size of $32$ and so on. Thus, we only enumerate the altered or newly added hyper-parameters in Table \ref{parmas}. As the window size is an obligatory parameter for the compared sparse-based methods, we set this by the mean degrees of the gene network graphs in each dataset for BigBird~\cite{zaheer_big_2020} and Diffuser~\cite{feng_diffuser_2022}, by $64$ for Longformer~\cite{beltagy_longformer_2020}.

\section{scGPT as Backbone: Varying the Sample Fraction}
Fig.\ref{fewshot_scgpt}(a)(b) illustrates the varying sample fraction experiments executed when integrating scGPT~\cite{cui_scgpt_2024} as backbones. Notably, when the fraction size is minimal, both versions of our model considerably enhance the performance of scGPT.

\section{Sensitivity Analysis Details: $K$}
Fig.\ref{fewshot_scgpt}(c) shows the detailed data distribution as the iteration steps, $K$, are varied. Generally, increasing $K$ significantly impacts the outcome when its value is minimal. However, when $K\geq6$, minor enhancements are observed, and a state of saturation seems to be attained. We adopt the most suitable $K$ for Pancreas datasets~\cite{chen_transformer_2023} and extend its use to other datasets. Nevertheless, given that different datasets might have diverse saturated points for $K$, it is advisable for users to either consider larger $K$ values or conduct additional trials when utilizing our model on their datasets.
\clearpage
\vspace{12pt}

\end{document}